\documentclass[12pt,a4]{article}

\usepackage{color,tikz}
\usepackage[unicode,bookmarks,bookmarksopen,bookmarksopenlevel=2,colorlinks,linkcolor=blue,citecolor=green]{hyperref}

\usepackage{amsmath,eucal,amssymb,amsthm,amsfonts}
\usepackage{mathrsfs,graphicx,texdraw}

\newtheorem{example}{Example}
\newtheorem{remark}{Remark}
\DeclareMathOperator{\tr}{tr}

\DeclareMathOperator{\diag}{diag}

\def\im{{\mbox{Im}}}

\def\openone{\leavevmode\hbox{\small1\kern-3.3pt\normalsize1}}

\def\re{\mathrm{Re\,}}
\def\const{\mathrm{const\,}}

\def\diag{\mbox{diag}\,}
\def\tr{\mbox{tr}\,}

\begin{document}

\begin{center}
{\LARGE \bf On the $N$-wave Equations with ${\cal PT}$-symmetry}

\bigskip

{\bf Vladimir S. Gerdjikov$^{a,}$\footnote{E-mail: {\tt gerjikov@inrne.bas.bg}},
Georgi  G. Grahovski$^{a,b,}$\footnote{E-mail: {\tt grah@essex.ac.uk}} and
Rossen  I. Ivanov$^{c,}$\footnote{E-mail: {\tt rivanov@dit.ie}}}

\end{center}

\medskip

\noindent
{\it $^{a}$ Institute for Nuclear Research and Nuclear Energy,
Bulgarian Academy of Sciences, 72 Tsarigradsko chaussee, 1784 Sofia, BULGARIA }\\
{\it $^{b}$ Department of Mathematical Sciences, University of Essex, Wivenhoe Park, Colchester CO4 3SQ, UK}\\
{\it $^{c}$ School of Mathematical Sciences, Dublin Institute of Technology, Kevin Street, Dublin 8, IRELAND }

\begin{abstract}
\noindent
We study extensions of $N$-wave systems with ${\cal PT}$-symmetry. The types of
(nonlocal) reductions leading to integrable equations invariant with respect to ${\cal P}$- (spatial reflection)
and ${\cal T}$- (time reversal) symmetries is described. The corresponding constraints on the fundamental
analytic solutions and the scattering data are derived. Based on examples of $3$-wave (related to the algebra $sl(3,{\Bbb C})$)
and $4$-wave (related to the algebra $so(5,{\Bbb C})$) systems, the properties of different types of 1- and 2-soliton solutions
are discussed. It is shown that  the $\mathcal{PT}$ symmetric $3$-wave equations may have
 regular multi-soliton solutions for some specific choices of their parameters.
\end{abstract}

\tableofcontents

\section{Introduction}\label{sec:1}

The $N$-wave equations \cite{ZM,ZMNP,ZaSh*74a}
\begin{equation}\label{eq:Nw}\begin{split}
i[J,Q_t] - i[I,Q_x] +[[I,Q],[J,Q]] = 0,
\end{split}\end{equation}
appear to be one of the important classes of integrable nonlinear evolution
equations (NLEE) with various applications in physics (nonlinear optics, plasma physics, fluid mechanics, quantum filed theory,
to name but a few). Generically, $Q(x,t)$ is assumed to be a $n\times n$ traceless matrix with vanishing diagonal entries and $J$ and $I$ are  constant diagonal matrices:
\begin{equation}\label{eq:JK}\begin{split}
J = \diag (a_1, a_2, \dots , a_n), \qquad  I= \diag (b_1,b_2, \dots , b_n), \qquad \tr J = \tr I =0.
\end{split}\end{equation}
Here also the subscripts $x$ and $t$ stay for differentiation with respect to $x$ and $t$. Typically $Q(x,t)$ can be chosen to be hermitian $Q(x,t) = Q^\dag (x,t)$.

The integrability of the $N$-wave system is based on its Lax representation discovered by Zakharov and Manakov
\cite{ZM,ZMNP}. It can be  formulated as the  condition $[L, M]=0$  of two ordinary differential operators $L$ and $M$
\begin{equation}\label{eq:Lax}\begin{split}
L\psi &\equiv i \frac{\partial \psi}{ \partial x } + U(x,t,\lambda)\psi(x,t,\lambda)=0, \qquad U(x,t,\lambda)= [J,Q](x,t)- \lambda J, \\
M\psi &\equiv i \frac{\partial \psi}{ \partial t } + V(x,t,\lambda)\psi(x,t,\lambda)=0, \qquad V(x,t,\lambda)= [I,Q](x,t)- \lambda I
\end{split}\end{equation}
commute identically with respect to the spectral parameter $\lambda$.
Then the  the $N$-wave equations can be solved using the inverse scattering method (ISM) \cite{ZM,ZMNP,FaTa,GVYa*08}: the associated spectral problem for the operator $L$ is
known as Zakharov-Shabat (ZS) type (matrix) spectral problem.

$N$-wave interaction models describe a special class of wave-wave
interactions  which are not sensitive on the physical nature of the waves and bear
an universal character. This explains why they find numerous applications in
physics and attract  attention of the scientific community over the last few
decades \cite{AckMil,Fer*95,67,58,vgrn,66,gk96,78,ISK*99,K,DJK,KRB,Za1*76}.

In the last decade artificial heterogenic media gained a special interest in Nonlinear Optics. This is related to the
fact that such media exhibit a number of properties that cannot be observed in homogeneous media. Such new properties,
 due to the resonance type of interaction of the media and light are observed in photonic crystals, random lasers,
 etc (for an up-to-date review, see \cite{UFN}). An important type of such models, allowing rigorous treatment are the so-called
 ${\cal PT}$-symmetric (Parity-Time) symmetric systems \cite{Konotop,Barash,Barash1,Bender1,Bender2, Ali1,Ali2,Nature}.

The initial interest in such systems was motivated by quantum mechanics \cite{Bender1, Ali1}. In \cite{Bender1}
it was shown that quantum systems with a non-hermitian Hamiltonian admit states with real eigenvalues, i.e. the hermiticity
of the Hamiltonian is not a necessary condition to have  real spectrum. Using such Hamiltonians one can build  up new quantum mechanics
\cite{Bender1,Bender2, Ali1,Ali2}.  Starting point is the fact that in the case of a non-Hermitian Hamiltonian with real spectrum,
the modulus of the wave function for the eigenstates is time-independent even in the case of complex potentials.

Historically the first pseudo-hermittian hamiltonian with real spectrum is the ${\cal PT}$-symmetric one in \cite{Bender1}. Pseudo-hermiticity here means that
the Hamiltonian ${\cal H}$ commutes with the operators of spatial reflection ${\cal P}$ and time reversal ${\cal T}$: ${\cal P}{\cal T}{\cal H}={\cal H}{\cal P}{\cal T}$.
The action of these operators is defined as follows: ${\cal P}: x\to -x$ and ${\cal T}: t\to -t$.

Due to Wigner's theorem, operators of symmetries can be either linear and unitary or anti-linear and anti-unitary. The type of
the operators ${\cal P}$ and ${\cal T}$ is determined by their action on the imaginary unit $i$: a direct calculation shows that ${\cal P}$ is
a linear and unitary operator. Since applying the spatial reflection twice will reproduce the initial state, so the wave functions $\psi(x,t)$ and ${\cal P}^2\psi(x,t)$
can differ only by a phase factor $\phi$: ${\cal P}^2 \psi(x,t)={\rm e}^{i\phi}\psi(x,t)$. In order to have the parity of the wave function to be observable quantity,
the operator ${\cal P}$ must be Hermitian. The only phase factor keeping the operator ${\cal P}$ Hermitian is the trivial one:  ${\cal P}^2 \psi(x,t)=\psi(x,t)$.
Supposing that the wave fuction is a scalar, this leads to the following action of the operator of spatial reflection on the space of states:
\begin{equation}\label{eq:P}
{\cal P}\psi(x,t)=\psi(-x,t).
\end{equation}
Applying similar arguments to the (anti-linear and anti-unitary) time reversal operator ${\cal T}$ show that it should act on the space of states as follows:
\begin{equation}\label{eq:T}
{\cal T}\psi(x,t)=\psi^*(x,-t).
\end{equation}
Therefore, the Hamiltonian and the wave function are ${\cal PT}$-symmetric, if
\begin{equation}\label{eq:PT}
{\cal H}(x,t)={\cal H}^*(-x,-t), \qquad \psi(x,t)=\psi^*(-x,-t).
\end{equation}
Integrable systems with ${\cal PT}$-symmetry were studied extensively over the last two decades \cite{Fring1,Fring2}.
Recently, in \cite{AblMus,GeSa} was proposed  a nonlocal integrable equation of nonlinear Schr\"odinger type with
${\cal PT}$-symmetry, due to the invariance of
the so-called self-induced potential $V(x,t)=\psi(x,t)\psi^*(-x,-t)$ under the combined action of parity and time
reversal symmetry (\ref{eq:PT}). In the same paper, the 1-soliton solution for this model is derived and it was shown
that it develops  singularities in finite time. Soon after this, nonlocal ${\cal PT}$-symmetric generalisations are found for
 the Ablowitz-Ladik model in \cite{AblMus1}. All these models are integrable by the Inverse Scattering Method (ISM) \cite{AblMus2}.

The purpose of the present paper is to study ${\cal PT}$-symmetric extensions of the $N$-wave equations: in what follows we
will outline the $\mathcal{PT}$-reductions for  $3$- and $4$-wave systems (with Lax operators related to the algebras $sl(3,{\Bbb C})$
and $so(5,{\Bbb C})$ respectively) and will derive the constraints these reductions impose on the scattering data of the Lax operators. We will derive their 1- and 2-soliton solutions.

The paper is organized as follows. The next Section 2 contains preliminaries on the Mikhailov reduction group.
In the next Section 3 we give some basic examples of three- and four-wave interactions.
Here we also formulate the reduction conditions compatible with the $\mathcal{PT}$-symmetry. In Section 4
we construct the FAS of the Lax operators and derive the constraints that the  $\mathcal{PT}$-symmetries
impose on them, on the scattering matrix $T(\lambda,t)$ and on the Gauss factors of $T(\lambda,t)$.
We also discuss the consequences of the reductions on the locations of the discrete eigenvalues of $L$.
In Section 5 we derive and analyze the soliton solutions of 3-wave systems related to the algebra
$sl(3,{\Bbb C})$.

\section{Preliminaries}

Here we outline basic facts about the direct scattering problem for the Lax operator (\ref{eq:Lax}). We outline
also the main types of local and nonlocal reductions of (\ref{eq:Lax}). The reduction problem \cite{2} plays a fundamental
role one in the theory of integrable systems. The class of nonlocal reductions  includes  ${\cal P}$- (spatial reflection) and
${\cal T}$-(time reversal) symmetries \cite{AblMus,AblMus2,TV}.

\subsection{Direct scattering transform and Jost solutions}\label{ssec:2.2}

The starting point here are  the so-called Jost solutions for the operator $L(\lambda)$,  which are defined by their asymptotics (see, e.g. \cite{GVYa*08}
and the references therein):
\begin{equation}\label{eq:Jost}
\lim_{x \to -\infty} \phi(x,t,\lambda) e^{  i \lambda J x }=\openone, \qquad  \lim_{x \to \infty}\psi(x,t,\lambda) e^{  i\lambda J x } = \openone
 \end{equation}
and the scattering matrix $T(\lambda)$ is defined by $T(\lambda,t)\equiv
\psi^{-1}\phi(x,t,\lambda)$. Here we assume that the potential $q(x,t)$ is tending to zero fast enough, when $|x|\to \infty$.
We note also, that since  $U(x,t,\lambda)$ and $V(x,t,\lambda)$ are taking values in simple Lie algebras ($sl(3)$ and
$so(5)$ as regards our examples), then the Jost solutions and the scattering matrix $T(t,\lambda)$ must  take
values in the corresponding  Lie group.

An  essential difficulty that comes up here is due to the fact that the Jost solutions $\psi(x,t,\lambda)$ and $\phi(x,t,\lambda)$
exist only if the potential $Q(x,t)$ is on finite support \cite{BeCo}. In this case not only the Jost solutions, but also the scattering matrix $T(\lambda,t)$
 (here and hereafter by $\hat{X}$ we will denote the inverse matrix, $\hat{X}\equiv X^{-1}$):
\begin{equation}\label{eq:TG}\begin{aligned}
T(t,\lambda) = \hat{\psi}(x,t,\lambda) \phi(x,t,\lambda).
\end{aligned}\end{equation}
become meromorphic functions of $\lambda$. More details and more precise statements will be given in the next Section 3.

\subsection{Local and Non-local Reductions}\label{ssec:2.3}

An important and systematic tool to construct new integrable NLEE is the so-called  reduction group \cite{2}.
It will be instructive to start with the local reductions:
\begin{equation}\label{eq:Z2-Mi}\begin{aligned}
& \mbox{1)}& \qquad C_1(U^{\dagger}(\kappa _1(\lambda ))) &= U(\lambda ),
&\quad C_1(V^{\dagger}(\kappa _1(\lambda ))) &= V(\lambda ), \\
 & \mbox{2)} & \qquad C_2(U^{T}(\kappa _2(\lambda ))) &= -U(\lambda ), &\quad
C_2(V^{T}(\kappa _2(\lambda ))) &= -V(\lambda ), \\
& \mbox{3)}& \qquad C_3(U^{*}(\kappa _1(\lambda ))) &= -U(\lambda ), &\quad
C_3(V^{*}(\kappa _1(\lambda ))) &= -V(\lambda ), \\
 &\mbox{4)}& \qquad C_4(U(\kappa _2(\lambda ))) &= U(\lambda ), &\quad
C_4(V(\kappa _2(\lambda ))) &= V(\lambda ).
\end{aligned}\end{equation}
The consequences of these reductions and the constraints they impose on the FAS and the Gauss
factors of the scattering matrix are well known, see \cite{2,ZMNP,67,58,vgrn,78}.

It is important to note, that for the $N$-wave equations {\em there are no }  reductions  compatible with either $\mathcal{P}$- or  $\mathcal{T}$-symmetry
separately. However the ${\Bbb Z}_2$ reductions
\begin{equation}\label{eq:Z2-nl}\begin{aligned}
& \mbox{1)}& \;\; C_1(U^{\dagger}(-x,-t,\kappa _1\lambda^* )) &= -U(x,t,\lambda ),
&\;C_1(V^{\dagger}(-x,-t,\kappa _1\lambda^* )) &= -V(x,t,\lambda ), \\
 & \mbox{2)} & \;\; C_2(U^{T}(-x,-t,\kappa _2\lambda )) &= U(x,t,\lambda ), &\;C_2(V^{T}(-x,-t, \kappa _2\lambda )) &= V(x,t,\lambda ), \\
& \mbox{3)}& \;\; C_3(U^{*}(-x,-t,\kappa _1\lambda^*)) &= U(x,t,\lambda ), &\;C_3(V^{*}(-x,-t,\kappa _1\lambda^*)) &= V(x,t,\lambda ), \\
 &\mbox{4)}& \;\; C_4(U(-x,-t,\kappa _2\lambda )) &= -U(x,t,\lambda ), &\;C_4(V(-x,-t,\kappa _2\lambda )) &= -V(x,t,\lambda ),
\end{aligned}\end{equation}
are obviously ${\cal PT}$-symmetric \cite{TV}. Here $\kappa_i^2 =1$ and $C_i$, $i=1,\dots, 4$ are involutive automorphisms of the relevant Lie algebra.
Below we will present examples of such reductions for $3$-wave and $4$-wave interaction systems;
it is natural to call them {\it non-local}.

\section{Basic examples of $3$-wave and $4$-wave interactions}\label{ssec:2.4}

\subsection{Examples of $3$-wave Interactions}

We consider a Lax pair of the form (\ref{eq:Lax}) related to $sl(3, {\Bbb C})$ algebra: in the typical representation the potential matrix adopts the form
\begin{equation}\label{eq:*2}
Q = \begin{pmatrix}  0 & q_1(x,t) & q_3(x,t) \\  p_1(x,t) & 0 & q_2(x,t) \\
p_3(x,t) & p_2(x,t) & 0 \end{pmatrix},
\qquad \begin{aligned}
J &= \diag ( a_1, a_2 , a_3) , \\ I &= \diag ( b_1 , b_2 , b_3).
\end{aligned}\end{equation}
We presume that the potential matrices $Q(x,t)$, $I$ and $J$ are traceless (i.e. the Lax operators take values
in the algebra $sl(3, {\Bbb C})$) and the eigenvalues of $I$ and $J$ are ordered
as follows: $a_1>a_2>a_3$, $b_1>b_2>b_3$ ($a_1+a_2+a_3=0$ and $b_1+b_2+b_3=0$).

The compatibility condition (\ref{eq:3w}) produces the generic system of 3-wave resonant interaction equations
\begin{equation}\label{eq:3w}\begin{split}
&i(a_1-a_2) q_{1,t} -i(b_1-b_2) q_{1,x} + \kappa  q_3(x,t)p_2(x,t) =0, \\
& i(a_2-a_3) q_{2,t} -i(b_2-b_3) q_{2,x} + \kappa q_3(x,t) p_1(x,t) =0, \\
& i(a_1-a_3) q_{3,t} -i(b_1-b_3) q_{3,x} + \kappa q_1(x,t) q_2(x,t) =0,
\end{split}\end{equation}
where the interaction constant $\kappa$ is
\begin{equation}\label{eq:kapa}\begin{split}
  \kappa = a_1(b_2 - b_3) - a_2(b_1 - b_3) + a_3(b_1 - b_2).
\end{split}\end{equation}
In addition there are 3  ``conjugated'' equations, formally obtained from (\ref{eq:3w}) by replacing
$p_k (x,t)$ by $q_k^* (x,t)$. The standard reduction $p_k (x,t)=q_k^* (x,t)$  ($k=1\div 3$) leads to
the canonical form of the 3-wave resonant interaction equations \cite{ZM,ZMNP}.

The system (\ref{eq:3w}) is Hamiltonian \cite{ZM,ZMNP} and possess a hierarchy of pairwise compatible
Hamiltonian structures \cite{G*86,GVYa*08}.
The (canonical) Hamiltonian  of (\ref{eq:Nw}) is given by \cite{ZM,ZMNP,vgrn,K,KRB}:
\begin{eqnarray}\label{eq:1.5}
H_{\rm 3-w} ={1 \over 2}\int_{-\infty }^{\infty }  dx\, \left(  \sum_{k=1}^{3} v_k\left(  q_k \frac{\partial p_k}{ \partial x }
- p_k \frac{\partial q_k}{ \partial x }\right) +\kappa (q_3 p_1 p_2 - p_3q_1q_2 ) \right).
\end{eqnarray}
Here $v_k$ are the group velocities of  (\ref{eq:3w}):
\[
v_1 = {b_1-b_2\over a_1-a_2}, \qquad v_2 = {b_2-b_3\over a_2-a_3}, \qquad v_3 = {b_1-b_3\over a_1-a_3}.
\]

\begin{example}\label{exa:1A}
 If we impose a reduction of type 1) from (\ref{eq:Z2-nl}) with $C_1={\rm id}$, we will get
$a_{k}^*=\epsilon a_{k}$, $b_{k}^*=\epsilon b_{k}$ and $p_k(x,t)=-\epsilon q_k^*(-x,-t)$ ($k=1\div 3$).
Thus the generic 3-wave system (\ref{eq:3w}) will reduce to following nonlocal one
\begin{equation}\label{eq:3wa}\begin{split}
&i(a_1-a_2) q_{1,t} -i(b_1-b_2) q_{1,x} -\epsilon \kappa  q_3(x,t)q_2^*(-x,-t) =0, \\
& i(a_2-a_3) q_{2,t} -i(b_2-b_3) q_{2,x} -\epsilon \kappa q_3(x,t) q_1^*(-x,-t) =0, \\
& i(a_1-a_3) q_{3,t} -i(b_1-b_3) q_{3,x} + \kappa q_1(x,t) q_2(x,t) =0.
\end{split}\end{equation}

\end{example}

\begin{example}\label{exa:1C}
If we impose a reduction of type 1) from (\ref{eq:Z2-nl}) with
\[
C_1=\left(\begin{array}{ccc} 0 & 0 & 1\\ 0& 1 & 0\\ 1 & 0 & 0
\end{array} \right)
\]
will give
\begin{eqnarray}\label{eq:a2-r3}
&&a_{3}^*=\epsilon a_{1}, \qquad a_{2}^*=\epsilon a_{2}, \qquad b_{3}^*=\epsilon b_{1}, \qquad b_{2}^*=\epsilon b_{2},\\
&& p_1(x,t)=-\epsilon q_2^*(-x,-t), \qquad p_2(x,t)=-\epsilon q_1^*(-x,-t), \qquad p_3(x,t)=-\epsilon q_3^*(-x,-t).\nonumber
\end{eqnarray}
As a result, one can get the following nonlocal 3-wave system:
\begin{equation}\label{eq:3w-c}\begin{split}
&i(a_1-a_2) q_{1,t} -i(b_1-b_2) q_{1,x} - \kappa  q_3(x,t)q_1^*(-x,-t) =0, \\
& i(a_2-a_3) q_{2,t} -i(b_2-b_3) q_{2,x} - \kappa q_3(x,t) q_2(-x,-t) =0, \\
& i(a_1-a_3) q_{3,t} -i(b_1-b_3) q_{3,x} + \kappa q_1(x,t) q_2(x,t) =0.
\end{split}\end{equation}
\hfill $\Box$

\end{example}

\subsection{Examples of $4$-wave interactions}

Another important example  are the $4$-wave equations. Their Lax operators are related to the algebra $so(5, {\Bbb C})$. In the
typical representation of $so(5, {\Bbb C})$ the (generic) potential $Q(x,t)$ takes the form:
\begin{eqnarray}\label{grah:ex-4w}
Q(x,t) =
\left(\begin{array}{ccccc} 0 & q_{1}(x,t) & q_{3}(x,t) & q_{4}(x,t) & 0\\
p_{1}(x,t) &0 & q_2(x,t) & 0 & q_{4}(x,t) \\ p_{3}(x,t) & p_2(x,t) & 0 & q_2(x,t) & -q_{3}(x,t)\\
p_{1}(x,t) & 0 & p_2(x,t) & 0 & q_{1}(x,t)\\ 0 & p_{4}(x,t) & -p_{3}(x,t) & p_{1}(x,t) & 0
\end{array} \right);
\end{eqnarray}
and the elements of the Cartan subalgebra $J$ and $I$ can be represented as diagonal matrices
\begin{equation}\label{eq:JK2}\begin{split}
J=\diag(a_1, a_2, 0 ,-a_2, -a_1), \qquad I=\diag(b_1, b_2, 0 ,-b_2, -b_1).
\end{split}\end{equation}
The zero-curvature representation (\ref{eq:Nw}) will lead to the following system of 4 waves  \cite{vgrn}:
\begin{equation}\label{eq:b2-4w}\begin{aligned}
& i(a_1-a_2)q_{1,t}-i(b_1-b_2)q_{1,x}+2\kappa q_{3}(x,t)p_{2}(x,t)=0, \\
&ia_2q_{2,t}-ib_2q_{2,x}+\kappa [q_{4}(x,t)p_3(x,t)+q_{3}(x,t)p_{1}(x,t)]=0,\\
&ia_1q_{3,t}-ib_1q_{3,x}+\kappa [q_{4}(x,t)p_{2}(x,t)-q_{1}(x,t)q_{2}(x,t)]=0, \\
&i(a_1+a_2)q_{4,t}-i(b_1+b_2)q_{4,x}-2\kappa q_{3}(x,t)q_{2}(x,t)=0,
\end{aligned}\end{equation}
where the interaction constant reads $\kappa =a_1b_2-a_2b_1 $. In addition we have a set of 4 ``conjugated''
equations, formally obtained from (\ref{eq:b2-4w}) by replacing $p_k (x,t)$ by $q_k^* (x,t)$. The standard
reduction $p_k (x,t)=q_k^* (x,t)$  ($k=1\div 4$) leads to the canonical form of the 4-wave resonant interaction equations \cite{ZMNP,58,vgrn}.
Here we have a local reduction of type 1).

\begin{example}\label{exa:2A}
Let us impose a reduction of type 3) from (\ref{eq:Z2-nl}) with
 \begin{equation}\label{eq:C1}\begin{split}
  C_1=\left(\begin{array}{ccccc} 0 & 1 & 0 & 0 & 0\\
1 &0 & 0 & 0 & 0 \\ 0 & 0 & 1 & 0 & 0\\ 0 & 0 & 0 & 0 & -1\\ 0 & 0 & 0 & -1 & 0
\end{array} \right),
 \end{split}\end{equation}
This  gives  $C_1J^*C_1 = J$ and $C_1 Q^*(-x,-t) C_1^{-1} =Q$; or in components
we have $a_2=a_1^*$ and
\begin{equation}\label{eq:b2-ex3}\begin{aligned}
p_1(x,t) &= q_1^*(-x,-t), &\quad q_2(x,t) &= q_3^*(-x,-t),  &\quad p_2(x,t)= p_3^*(-x,-t),\\
q_4(x,t) &= -q_4^*(-x,-t),  &\quad p_4(x,t) &=- p_4^*(-x,-t).
\end{aligned}\end{equation}
 Thus we find the following nonlocal analogue of the classical 4-wave  equations
\begin{equation}\label{eq:b2-4w-nl}\begin{split}
& i(a_1-a_2)q_{1,t}-i(b_1-b_2)q_{1,x}+  \kappa q_{3}(x,t)p_{3}^*(-x,-t)=0,\\
&ia_2q_{2,t}-ib_2q_{2,x}+\kappa [q_{4}(x,t)p_2^*(-x,-t)+q_{1}^*(-x,-t)q_{1}^*(-x,-t)]=0,\\
&ia_2p_{2,t}-ib_2p_{2,x}-\kappa [p_{4}(x,t)q_{2}^*(-x,-t)+q_{1}(x,t)p_{2}^*(-x,-t)]=0, \\
&i(a_1+a_2)q_{4,t}-i(b_1+b_2)q_{4,x}-\kappa q_{2}(x,t)q_{2}^*(-x,-t)=0.\\
&i(a_1+a_2)p_{4,t}-i(b_1+b_2)p_{4,x}+\kappa p_{2}(x,t)p_{2}^*(-x,-t)=0.
\end{split}\end{equation}
 \end{example}

\begin{example}\label{exa:2B}
If we impose a reduction of type 3) from (\ref{eq:Z2-nl}) with $C_3={\rm id}$ is equivalent to
imposing a reduction of type 1) with $C_1=w_0$, where $w_0$ is the (inner) automorphism which maps the
highest-weight vectors into the lowest-weight vectors of the algebra $so(5, {\Bbb C})$ (see e.g. \cite{58}). This will lead to:
$a_{1,2}^*=-\epsilon a_{1,2}$, $b_{1,2}^*=-\epsilon b_{1,2}$, $q_k(x,t)=\epsilon q_k^*(-x,-t)$ and
$p_k(x,t)=\epsilon p_k^*(-x,-t)$  ($k=1\div 4$). This will lead to a generic 4-wave system (for 8
real-valued functions $q_k(x,t)$ and $p_k(x,t)$, $k=1\div 4$) of the type (\ref{eq:b2-4w}).

\end{example}

\begin{example}\label{exa:2C}
 If we impose a reduction of type 1) from (\ref{eq:Z2-nl}) with $C_1$ given by (\ref{eq:C1}) above
we will get:
$a_{2}^*= a_{1}$, $b_{2}^*= b_{1}$ and
\begin{equation}\label{eq:b2-r3}\begin{aligned}
q_1(x,t) &= q_1^*(-x,-t), &\quad p_2(x,t) &= q_3^*(-x,-t),  &\quad p_3(x,t)= q_2^*(-x,-t),\\
 p_1(x,t) &= p_1^*(-x,-t),  &\quad p_4(x,t) &=- q_4^*(-x,-t).
\end{aligned}\end{equation}
As a result, one gets the  5-wave system (containing 2 ``real'' ($p_1$ and $q_1$) and 3 complex ($q_2$, $q_3$ and $q_4$) functions):
\begin{equation}\label{eq:b2-4wp}\begin{split}
& i(a_1-a_2)q_{1,t}-i(b_1-b_2)q_{1,x}+ \kappa q_{3}(x,t)q_{3}^*(-x,-t)=0,\\
& i(a_1-a_2)p_{1,t}-i(b_1-b_2)p_{1,x}- \kappa q_{2}(x,t)q_{2}^*(-x,-t)=0,\\
&ia_2q_{2,t}-ib_2q_{2,x}+\kappa [q_4(x,t)q_{2}^*(-x,-t)+q_{3}(x,t)p_{1}(x,t)]=0,\\
&ia_1q_{3,t}-ib_1q_{3,x}+\kappa [ q_{3}^*(-x,-t)q_{4}(x,t)-q_{1}(x,t)q_{2}(x,t)]=0, \\
& i(a_1+a_2)q_{4,t}-i(b_1+b_2)q_{4,x}- \kappa q_{2}(x,t)q_{3}(x,t)=0,
\end{split}\end{equation}
This is an example of reduction that imposes a symmetry condition directly on the solution profiles of the ``real'' solutions.
\hfill $\Box$

\end{example}

All these examples and their reductions allow a Hamiltonian formulation. The reduction conditions will
impose restrictions on the corresponding Hamiltonians, symplectic forms and integrals of motion \cite{58}.

\section{The FAS and their reductions}
In constructing the FAS of the relevant Lax operator $L$ we have to consider separately the cases
of real and complex-valued $J$.

\subsection{The case of real $J \in sl(3)$}
The important notion of FAS for generic $n\times n$ Zakharov-Shabat system with real-valued $J$ (\ref{eq:Nw}) was introduced by
Shabat \cite{Sh}. If $J$ has real eigenvalues then the Jost solutions of $L$ exist for all real $\lambda$.
They can be viewed as the solutions of a system of Volterra-type integral equations.
Indeed, if we introduce
\begin{equation}\label{eq:JostY}\begin{split}
Y_+(x,t,\lambda) =  \psi(x,t,\lambda) e^{iJ\lambda x}, \qquad  Y_-(x,t,\lambda) =  \phi(x,t,\lambda) e^{iJ\lambda x},
\end{split}\end{equation}
then $Y_\pm (x,t,\lambda)$ must satisfy:
\begin{equation}\label{eq:Y}\begin{split}
Y_{\pm; jk} (x,t,\lambda) = \delta _{jk} + i \int_{\pm \infty}^{x} dy\; e^{-i\lambda (a_j-a_k)(x-y)} \left(
[J,Q(y,t)]Y_\pm (x,t,\lambda)\right)_{jk}.
\end{split}\end{equation}
We will assume that the eigenvalues of $J$ are ordered, i.e. $a_1 >a_2 > \cdots >a_n$ and also
$\tr J=0$.

The Volterra equations (\ref{eq:JostY}) always have solution for real $\lambda$. Analytic extension
for $\lambda\in \mathbb{C}_+$ (resp. for $\lambda\in \mathbb{C}_-$) is possible only for the first column
of $Y_- (x,t,\lambda)$ and for the last column of $Y_+ (x,t,\lambda)$ (resp. for the last column
of $Y_- (x,t,\lambda)$ and for the first column of $Y_+ (x,t,\lambda)$. Shabat \cite{Sh} proposed to change the integral
equations (\ref{eq:JostY}) and consider two sets of integral equations:
\begin{equation}\label{eq:xi-p}\begin{split}
\xi^+_{jk} (x,t,\lambda) = \delta _{jk} + i \int_{\epsilon_{jk} \infty}^{x} dy\; e^{-i\lambda (a_j-a_k)(x-y)} \left(
[J,Q(y,t)]\xi^+ (y,t,\lambda)\right)_{jk}.
\end{split}\end{equation}
and
\begin{equation}\label{eq:xi-m}\begin{split}
\xi^-_{jk} (x,t,\lambda) = \delta _{jk} + i \int_{-\eta_{jk} \infty}^{x} dy\; e^{-i\lambda (a_j-a_k)(x-y)} \left(
[J,Q(y,t)]\xi^- (y,t,\lambda)\right)_{jk}.
\end{split}\end{equation}
where
\[ \epsilon_{jk}=  \begin{cases} 1 &\mbox{for} j< k,  \\ -1 &\mbox{for} j \geq k, \end{cases}, \qquad
 \eta_{jk}=  \begin{cases} -1 &\mbox{for} j\leq  k,  \\ 1 &\mbox{for} j > k, \end{cases}, \]
Then, in \cite{Sh}  he proved that the equations (\ref{eq:xi-p})  (resp. (\ref{eq:xi-m})) possess solutions $\xi^+ (x,t,\lambda) $
(resp. $\xi^- (x,t,\lambda) $) which allow analytic extension for $\lambda \in \mathbb{C}_+$
(resp. for $\lambda \in \mathbb{C}_-$).
The solutions $\xi^\pm (x,t,\lambda) $ can be viewed also as solutions to a Riemann-Hilbert problem (RHP) in multiplicative form \cite{67,Sh,DokLeb,ZMNP}
\begin{equation}\label{eq:xi}\begin{split}
 \xi^+ (x,t,\lambda) = \xi^- (x,t,\lambda) G(x,t,\lambda), \qquad  G(x,t,\lambda) = e^{i\lambda Jx} G_0(t,\lambda)
 e^{-i\lambda Jx}
\end{split}\end{equation}
with canonical normalization, i.e. $\lim_{\lambda \to\infty} \xi^\pm (x,t,\lambda) =\openone$.

If we denote by $ \chi^\pm (x,t,\lambda) = \xi^\pm (x,t,\lambda) e^{-i\lambda Jx}$ then $\chi^\pm (x,t,\lambda) $
will be the FAS of $L$. Besides it was established \cite{IP2,ZMNP}, that the FAS are related to the Jost solutions by
\begin{equation}\label{eq:chi-pm}\begin{aligned}
\chi^\pm (x,t,\lambda) &= \phi(x,t,\lambda) S^\pm (t,\lambda), &\qquad  \chi^\pm (x,t,\lambda) &= \psi(x,t,\lambda) T^\mp (t,\lambda) D^\pm (\lambda),
\end{aligned}\end{equation}
where $S^\pm$, $T^\pm$ and $D^\pm$ are the Gauss factors of the scattering matrix:
\begin{equation}\label{eq:ScatM}\begin{aligned}
T(t,\lambda) =  T^-(t,\lambda) D^+(\lambda) \hat{S}^+(t,\lambda) =   T^+(t,\lambda) D^-(\lambda) \hat{S}^-(t,\lambda).
\end{aligned}\end{equation}
In other words $ T^-(t,\lambda)$, $ S^-(t,\lambda)$  (resp. $ T^+(t,\lambda)$, $ S^+(t,\lambda)$) are lower- (resp. upper-) triangular
matrices whose diagonal elements are equal to 1; the matrices $D^+(\lambda)$  and $D^-(\lambda)$ are diagonal and
allow analytic extension for $\lambda\in \mathbb{C}_\pm$ respectively. Then the sewing matrix $G_0(t,\lambda) =\hat{S}^-(t,\lambda)
S^+(t,\lambda)$. Along with $\chi^\pm (x,t,\lambda)$ one may use also $\tilde{\chi}^\pm (x,t,\lambda)$ defined by:
\begin{equation}\label{eq:chit-pm}\begin{split}
\tilde{\chi}^\pm (x,t,\lambda)  &= \phi(x,t,\lambda) S^\pm (t,\lambda) \hat{D}^\pm (\lambda), \qquad
\tilde{ \chi}^\pm (x,t,\lambda) = \psi(x,t,\lambda) T^\mp (t,\lambda).
\end{split}\end{equation}
i.e. $\tilde{\chi}^\pm (x,t,\lambda)=\chi^\pm (x,t,\lambda) \hat{D}^\pm (\lambda)$.

\subsection{The case of complex-valued $J \in sl(3)$}

We mentioned above the problem, that the Jost solutions $\psi(x,t,\lambda)$ and $\phi(x,t,\lambda)$
exist only if the potential $Q(x,t)$ is on finite support \cite{BeCo}. In this case not only the Jost solutions,
but also the scattering matrix $T(\lambda,t)$  and its factors of Gauss decompositions (\ref{eq:TG})
become meromorphic functions of $\lambda$.
However, following \cite{BeCo} and \cite{VYa} one can construct and prove the existence of fundamental analytic solutions
(FAS) $\chi_\nu(x,t,\lambda)$ of $L$. The construction is based on the set of  integral equations:
\begin{equation}\label{eq:Ieq1}\begin{split}
 \xi_{\nu;jk} (x,t,\lambda) &= \delta_{jk} + i \int_{\varepsilon_{\nu;jk}\infty }^{x} dy\; e^{-i \lambda (a_j-a_k)(x-y)} \left(
 [J,Q(y,t)] \xi_\nu(y,t,\lambda) \right)_{jk},
\end{split}\end{equation}
where the appropriately chosen $\varepsilon_{\nu; jk}$ take values $\pm 1$. It is obvious that for any choice of $\varepsilon_{\nu; jk}$
the solution of the equations (\ref{eq:Ieq1}) will be a solution also of the Lax operator.

Skipping the details, we will briefly outline the idea of constructing the FAS. The first step is to determine the regions (sectors) of the
analyticity of $\xi_\nu (x,t,\lambda)$. In each of these regions $\lambda \in \Omega_\nu$ the equations (\ref{eq:Ieq1}) will possess a solution provided
all the exponentials $e^{-i \lambda (a_j-a_k)(x-y)}$ are decaying for $x,y \to \pm \infty$. The regions $\Omega_\nu$ will be divided by lines
$l_\nu$, on which the exponentials $e^{-i \lambda (a_j-a_k)(x-y)}$ oscillate. In other words the lines $l_\nu$ consists of those points of $\lambda$,
for which
\begin{equation}\label{eq:l-nu1}\begin{split}
\im \,\lambda (a_j -a_k) =0.
\end{split}\end{equation}
Therefore we have to solve the set of algebraic equations (\ref{eq:l-nu1}) for all possible choices of $j \neq k$; the signs $\varepsilon_{\nu;kk}$
can be chosen arbitrarily. The solution of eq. (\ref{eq:l-nu1}) is therefore  provided by:
\begin{equation}\label{eq:l-nu2}\begin{split}
\arg \lambda = -\arg (a_j-a_k) = - \arctan \frac{\im (a_j-a_k) }{\re (a_j-a_k)} =-\beta_{jk} ;
\end{split}\end{equation}
in other words, each line $l_\nu$ is in fact a ray closing angle $-\beta_{jk}$ with the real $\lambda$-axis.

Depending on the size of the matrices $n\times n$ and on the choice for the eigenvalues $a_k$ of $J$ we may have
various sets of rays describing the spectrum of $L$. Note, that the number of rays is always even $2h$. Indeed,
if $l_\nu$ is one the rays closing angle $-\beta_{jk}$ with the real $\lambda$-axis, then the ray $l_{\nu+h}$
closing an angle $\pi -\beta_{jk}$ will also satisfy the equation (\ref{eq:l-nu2}) and will be a part of the
continuous spectrum of $L$. Below in Figure \ref{fig:1} we describe the spectrum of $3\times 3$ $L$-operators with special
choices for $J$, compatible with the reduction.

\begin{remark}\label{rem:1}
We will take special care that the operators $L$ and $M$ have the same (continuous) spectra. Obviously,
we can approach the operator $M$ in the same way as $L$ and construct its FAS. Obviously the FAS of $M$
will be analytic in sectors determined by the eigenvalues of $I$. In order that both $J$ and $I$ determine
the same sectors of analyticity $\Omega_\nu$ they should have (up to a common constant) the same sets of eigenvalues. However, if
$I=\const J$ then the interaction constant $\kappa =0$. Therefore below we choose $I=J^*$ which ensures the
isospectrality of $L$ and $M$ and ensures also that interaction constant  $\kappa \neq 0$.
\end{remark}

The next step consists in establishing the relations between the Jost solutions and the FAS $\chi_\nu(x,t,\lambda)$ in
each of the sectors $\Omega_\nu$.

In order to be more specific we now consider $3\times 3$ Lax operator and impose on it the reduction (\ref{eq:Z2-nl}), type 1).
It can be done in two ways:
\begin{equation}\label{eq:RAB}\begin{aligned}
& \mbox{A)} &\quad s_0 U^\dag (-x,-t,-\lambda^*) s_0^{-1} &=- U(x,t,\lambda), \\
  &\mbox{B)} &\quad s_0 U^\dag (-x,-t,\lambda^*) s_0^{-1} &=- U(x,t,\lambda).
\end{aligned}\end{equation}
where $s_0 = \left (\begin{array}{ccc} 0 & 0 & 1 \\ 0 & -1 & 0 \\ 1 & 0 & 0  \end{array}\right)$.
An immediate consequence of the reductions (\ref{eq:RAB}) is
\begin{equation}\label{eq:RABJ}\begin{aligned}
&\mbox{A)} & \quad s_0 J^\dag s_0^{-1} &= J, &\quad s_0 Q^\dag (-x,-t) s_0^{-1} &=Q(x,t), \\
&\mbox{B)} & \quad s_0 J^\dag s_0^{-1} &= -J, &\quad s_0 Q^\dag (-x,-t) s_0^{-1} &=-Q(x,t),
\end{aligned}\end{equation}
We will also need two sets of FAS for each sector, namely:
\begin{equation}\label{eq:chi-nu1}\begin{aligned}
\chi_{\nu} (x,t, \lambda) &= \phi(x,t,\lambda) S^+_\nu (\lambda,t)=   \psi(x,t,\lambda) T^-_\nu (\lambda,t) D^+_\nu (\lambda), &\quad \lambda &\in l_\nu^+ ,\\
\chi_{\nu} (x,t, \lambda) &= \phi(x,t,\lambda) S^-_{\nu+1} (\lambda,t)=   \psi(x,t,\lambda) T^+_{\nu+1} (\lambda,t) D^-_{\nu+1} (\lambda), &\quad \lambda &\in l_{\nu+1} ^-,
\end{aligned}\end{equation}
and
\begin{equation}\label{eq:chi-nu2}\begin{aligned}
\tilde{\chi}_{\nu} (x,t, \lambda) &= \phi(x,t,\lambda) S^+_\nu (\lambda,t) \hat{D}^+_\nu (\lambda)=   \psi(x,t,\lambda) T^-_\nu (\lambda,t) , &\quad \lambda &\in l_\nu^+ ,\\
\tilde{\chi}_{\nu} (x,t, \lambda) &= \phi(x,t,\lambda) S^-_{\nu+1} (\lambda,t)\hat{D}^-_{\nu+1} (\lambda)=   \psi(x,t,\lambda) T^+_{\nu+1} (\lambda,t) ,
&\quad \lambda &\in l_{\nu+1}^-,
\end{aligned}\end{equation}
where by $\lambda \in l_\nu^\pm$ we mean that $\lambda = \lambda' e^{\pm i0}$, where $\lambda'\in l_\nu$.
In order to explain the notations above we use the fact that \cite{VYa} to each of the rays $l_\nu$ one can relate a root $\alpha_\nu$
of the algebra $sl(3)$ and an $sl_\nu(2)$ subalgebra generated by $E_\alpha$, $E_{-\alpha}$ and $H_\alpha$, see Table~\ref{tab:0} below. Then the corresponding scattering
matrix $T_\nu(\lambda)$ evaluated on $l_\nu $ takes values in $sl_\nu(2)$ and the Gauss factors $S_\nu^\pm(\lambda)$,  $T_\nu^\pm(\lambda)$
and  $D_\nu^\pm(\lambda)$ are given by:
\begin{equation}\label{eq:STDpm}\begin{aligned}
S_\nu^\pm (\lambda) &= \exp \left( s_{\nu,\alpha_\nu}^\pm(\lambda) E_{\pm \alpha_\nu}\right), &\;
T_\nu^\pm (\lambda) &= \exp \left( t_{\nu,\alpha_\nu}^\pm (\lambda) E_{\pm \alpha_\nu}\right), &\;
D_\nu^\pm (\lambda) &= \exp \left( d_{\nu,\alpha_\nu}^\pm (\lambda) H_{\pm \alpha_\nu}\right),
\end{aligned}\end{equation}

\begin{table}
  \centering
  case A)\quad
  \begin{tabular}{|c|c|c|c|c|c|c|}
    \hline
Ray &    $l_0$ & $l_1$ & $l_2$ & $l_3$ & $l_4$ & $l_5$ \\ \hline
Root &    $-e_1+e_2$ &  $e_2-e_3$ &  $e_1-e_3$ &  $e_1-e_2$ &  $-e_2+e_3$ &  $-e_1+e_3$ \\
Angle & $\beta_{12}$    & $\pi/2$ & $\pi -\beta_{12}$ & $\pi +\beta_{12}$ & $-\pi/2$ & $-\beta_{12}$ \\
    \hline
  \end{tabular} \\[10pt]
  case B)\quad   \begin{tabular}{|c|c|c|c|c|c|c|}
    \hline
Ray &    $l_0$ & $l_1$ & $l_2$ & $l_3$ & $l_4$ & $l_5$ \\ \hline
Root &    $e_1-e_2$ &  $e_1-e_3$ &  $e_2-e_3$ &  $-e_1+e_2$ &  $-e_1+e_3$ &  $-e_2+e_3$ \\
Angle & $-\gamma_{12}$    & $0$ & $\gamma_{12}$ & $\pi -\gamma_{12}$ & $\pi$ & $\pi +\gamma_{12}$ \\
    \hline
  \end{tabular}
  \caption{The correspondence between the rays $l_\nu$, roots $\alpha_\nu$ and the angles that $l_\nu$ closes with the real axis
  for case A) (where $a_1=j_0+ij_1$ and $\beta_{12} =\arctan \frac{j_1}{3j_0}$) and case B)
   (where $a_1=j_0+ij_1$ and $\gamma_{12} =\arctan \frac{3j_1}{j_0}$).}\label{tab:0}
\end{table}

It is well known that the local reductions impose constraints on each of the Jost solutions $\psi(x,t,\lambda)$ and $\phi(x,t,\lambda)$ separately.

The nonlocal reductions we are studying relate the two Jost solutions as follows:
\begin{equation}\label{eq:JostR}\begin{aligned}
&\mbox{A)} & \quad s_0 \psi^\dag (-x,-t, -\lambda^*) s_0^{-1} &= \hat{\phi}(x,t,\lambda), &\quad
&\mbox{B)} & \quad s_0 \psi^\dag (-x,-t,\lambda^*) s_0^{-1} &= \hat{\phi}(x,t,\lambda) .
\end{aligned}\end{equation}
Our next task is to derive the constraints that the nonlocal reductions impose on the FAS. Before to go into this we
find how the mappings $\lambda \to -\lambda^*$ and $\lambda \to \lambda^*$ affect the the lines $l_\nu$ and the
sectors $\Omega_\nu$:
\begin{equation}\label{eq:la-la*}\begin{aligned}
&\mbox{A)} &\; & \lambda \leftrightarrow -\lambda^* &\quad  l_\nu^+ &\leftrightarrow l_{4-\nu}^- &\quad  \Omega_\nu &\leftrightarrow
\Omega_{3-\nu} , &\; &\mbox{B)} &\; & \lambda \leftrightarrow \lambda^* &\quad  l_\nu^+ &\leftrightarrow l_{8-\nu}^- &\quad  \Omega_\nu &\leftrightarrow
\Omega_{7-\nu} ,
\end{aligned}\end{equation}
where  $4-\nu$ and $3-\nu$ in case A) and $8-\nu$ and $7-\nu$ in case B) are evaluated modulo 6.


\begin{figure}
\begin{tikzpicture}

\draw [-,dashed,black](-4,0) -- (-2.5,0);
\draw [-,dashed,black](-1.5,0) -- (1.5,0);
\draw [->,dashed,black](2.5,0) -- (4,0);
\draw [->,black](0,-4) -- (0,4);

\draw [-,ultra thick, blue](0,-3.5) -- (0,3.5);
\draw [-,ultra thick, blue, rotate=30](-3.5,0) -- (3.5,0);
\draw [-,ultra thick, blue, rotate=150](-3.5,0) -- (3.5,0);
\draw [black] (3.5,3.5) node  {{\small $\lambda$}};

\draw [black] (3.5,3.5) circle (0.25);

\draw [black] (0.25,3.75) node  {{\small  $l_1$}};
\draw [black] (3.13,1.5) node  {{\small  $l_0$}};
\draw [black] (-3.13,1.5) node  {{\small  $l_2$}};
\draw [black] (-0.25,-3.75) node  {{\small  $l_4$}};
\draw [black] (-3.23,-1.5) node  {{\small  $l_3$}};
\draw [black] (3.23,-1.5) node  {{\small  $l_5$}};

\draw [blue] (1.25,2.17) node  {{\large  $\Omega_0$}};
\draw [blue] (-1.25,2.17) node  {{\large  $\Omega_1$}};
\draw [blue] (-2.0,0.0) node  {{\large  $\Omega_2$}};
\draw [blue] (-1.25,-2.17) node  {{\large  $\Omega_3$}};
\draw [blue] (1.25,-2.17) node  {{\large  $\Omega_4$}};
\draw [blue] (2.0,0.0) node  {{\large  $\Omega_5$}};

\draw [black] (0,-4.5) node  {{\large  (A)}};

\end{tikzpicture}
\begin{tikzpicture}

\draw [-,dashed,black](0,-4) -- (0,-2.5);
\draw [-,dashed,black](0,-1.5) -- (0,1.5);
\draw [->,dashed,black](0,2.5) -- (0,4);
\draw [->,black](-4,0) -- (4,0);

\draw [-,ultra thick, blue](-3.5,0) -- (3.5,0);
\draw [-,ultra thick, blue, rotate=40](-3.5,0) -- (3.5,0);
\draw [-,ultra thick, blue, rotate=140](-3.5,0) -- (3.5,0);
\draw [black] (3.5,3.5) node  {{\small $\lambda$}};

\draw [black] (3.5,3.5) circle (0.25);

\draw [black] (3.75,0.25) node  {{\small  $l_1$}};
\draw [black] (3.00,2.5) node  {{\small  $l_2$}};
\draw [black] (-3.00,2.5) node  {{\small  $l_3$}};
\draw [black] (-3.75,0.25) node  {{\small  $l_4$}};
\draw [black] (-3.00,-2.5) node  {{\small  $l_5$}};
\draw [black] (3.00,-2.5) node  {{\small  $l_0$}};

\draw [blue] (2.0,0.75) node  {{\large  $\Omega_1$}};
\draw [blue] (0,2) node  {{\large  $\Omega_2$}};
\draw [blue] (-2.0,0.75) node  {{\large  $\Omega_3$}};
\draw [blue] (-2.0,-0.75) node  {{\large  $\Omega_4$}};
\draw [blue] (0,-2) node  {{\large  $\Omega_5$}};
\draw [blue] (2.0,-0.75) node  {{\large  $\Omega_0$}};

\draw [black] (0,-4.5) node  {{\large  (B)}};

\end{tikzpicture}
\caption{The continuous spectra of a $3\times 3$ Lax operator with complex-valued $J$: case A -- left panel;
case B -- right panel subject to a  reduction of type
A)and  B) in (\ref{eq:RAB}). }
\label{fig:1}
\end{figure}
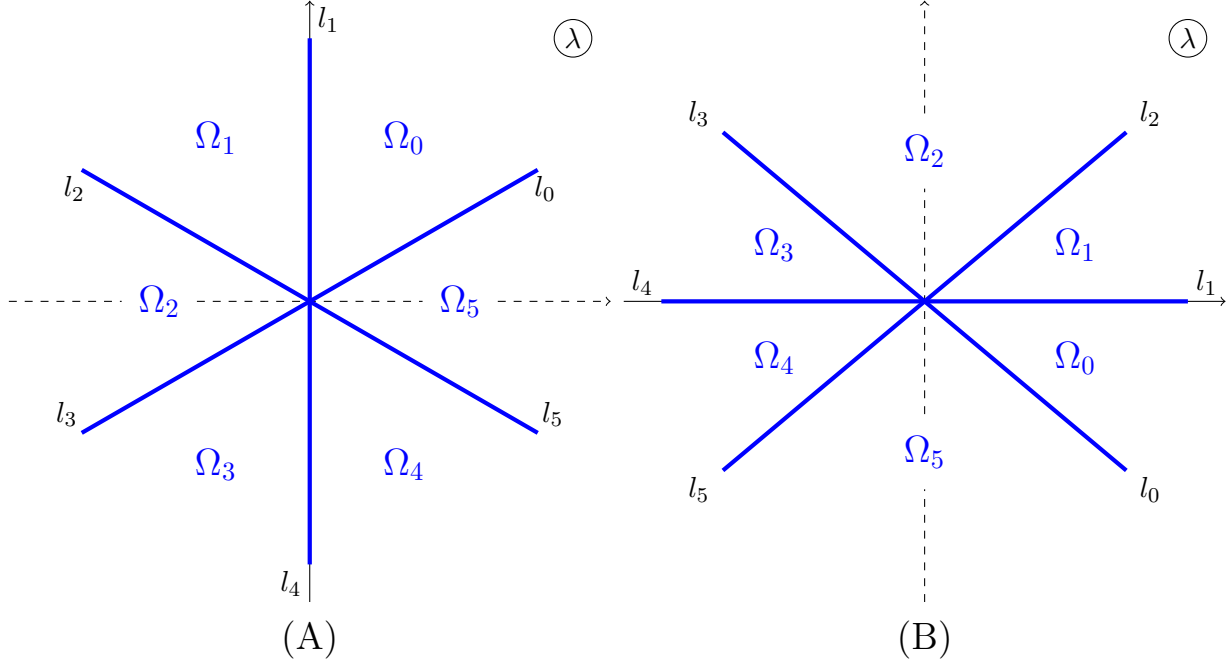

Next we compare the relations (\ref{eq:Jost}) with (\ref{eq:chi-nu1}) and (\ref{eq:chi-nu2}) fixing up
$\lambda$ to take values close to the rays $l_\nu$. This allows us to derive the constraints on the
scattering data.

In case A we find that:
\begin{equation}\label{eq:chi-redA}\begin{split}
s_0 (\chi_\nu (-x,-t,-\lambda^*))^\dag s_0^{-1} = \widehat{\tilde{\chi}}_{3-\nu} (x,t,\lambda),
\end{split}\end{equation}
where $3-\nu $ is evaluated modulo 6. From these relations there follows the constraints on the
Gauss factors of the scattering matrix. Skipping the tedious details we obtain;
\begin{equation}\label{eq:spm-red}\begin{aligned}
s_0 (S_\nu^+(-\lambda^*))^\dag s_0^{-1} &= \hat{T}_{4-\nu}^+(\lambda), &\;
s_0 (S_\nu^-(-\lambda^*))^\dag s_0^{-1} &= \hat{T}_{4-\nu}^-(\lambda), \\
s_0 (D_\nu^+(-\lambda^*))^\dag s_0^{-1} &= D_{4-\nu}^+(\lambda).
\end{aligned}\end{equation}
In case B we have
\begin{equation}\label{eq:chi-redB}\begin{split}
s_0 (\chi_\nu (-x,-t,-\lambda^*))^\dag s_0^{-1} = \widehat{\tilde{\chi}}_{7-\nu} (x,t,\lambda),
\end{split}\end{equation}
where $7-\nu $ is evaluated modulo 6. The constraints on the
Gauss factors of the scattering matrix now read;
\begin{equation}\label{eq:spm-redB}\begin{aligned}
s_0 (S_\nu^+(\lambda^*))^\dag s_0^{-1} &= \hat{T}_{8-\nu}^+(\lambda), &\;
s_0 (S_\nu^-(\lambda^*))^\dag s_0^{-1} &= \hat{T}_{8-\nu}^-(\lambda), &\;
s_0 (D_\nu^+(\lambda^*))^\dag s_0^{-1} &= D_{8-\nu}^+(\lambda),
\end{aligned}\end{equation}
and again $7-\nu$ and $8-\nu$ are evaluated modulo 6.

Let us also note, that the zeroes and the poles of the functions $D_\nu^\pm (\lambda)$ determine
the discrete eigenvalues of $L$. From eq. (\ref{eq:spm-red}) it follows that if $\lambda_k$ is a
discrete eigenvalue, then $-\lambda_k^*$ will also be an eigenvalue of $L$ in case A. In case B
eq. (\ref{eq:spm-redB})   ensures that if $\lambda_k$ is a discrete eigenvalue, then $\lambda_k^*$ will also be an eigenvalue of $L$.

\begin{remark}\label{rem:2}
The same pairing of the discrete eigenvalues $\lambda_k, -\lambda_k^*$ like in our case A was discovered in \cite{AblMus}
for the spectrum of the $\mathcal{P}$-symmetric $2\times 2$ Zakharov-Shabat  system. The substantial difference with our
case is that in \cite{AblMus} both $\lambda_k$ and $ -\lambda_k^*$ lie in the same analyticity region. From Figure \ref{fig:1}A
it is obvious the two eigenvalues always must lie in different analyticity regions.
\end{remark}

\subsection{The case of complex-valued $J \in so(5)$}

The root system $\Delta$ of $so(5)$ is a special set of 8 vectors in two-dimensional
Euclidean space $\mathbb{E}^2$. If we introduce in $\mathbb{E}^2$ the orthonormal basis $e_1, e_2$
then $\Delta \equiv \{ e_1 \pm e_2, -e_1 \pm e_2, \pm e_1, \pm e_2\}$. Any element $J$ from the  Cartan
subalgebra is a linear combination of two basic elements $J=a_1 H_{e_1}+a_2 H_{e_2}$. We can associate
with $J$ a vector $\vec{J} = a_1 e_1 + a_2 e_2$. For more details on the root system of $so(5)$ and other
simple Lie algebras \cite{Helg}.
The 4-wave system with real-valued $J$ and their soliton solutions have been analyzed in \cite{ZMNP,58,vgrn,RI,gc}.


\begin{figure}
\begin{center}
\begin{tikzpicture}

\draw [->,black](0,-4) -- (0,4);
\draw [->,black](-4,0) -- (4,0);

\draw [-,ultra thick, blue](-3.5,0) -- (3.5,0);
\draw [-,ultra thick, blue, rotate=90](-3.5,0) -- (3.5,0);
\draw [-,ultra thick, blue, rotate=40](-3.5,0) -- (3.5,0);
\draw [-,ultra thick, blue, rotate=140](-3.5,0) -- (3.5,0);
\draw [black] (3.5,3.5) node  {{\small $\lambda$}};

\draw [black] (3.5,3.5) circle (0.25);

\draw [black] (3.75,0.25) node  {{\small  $l_0$}};
\draw [black] (3.00,2.5) node  {{\small  $l_1$}};
\draw [black] (-3.00,2.5) node  {{\small  $l_3$}};
\draw [black] (-3.75,0.25) node  {{\small  $l_4$}};
\draw [black] (-3.00,-2.5) node  {{\small  $l_5$}};
\draw [black] (3.00,-2.5) node  {{\small  $l_7$}};
\draw [black] (0.25,3.75) node  {{\small  $l_2$}};
\draw [black] (0.25,-3.75) node  {{\small  $l_6$}};

\draw [blue] (2.0,0.75) node  {{\large  $\Omega_0$}};
\draw [blue] (-2.0,0.75) node  {{\large  $\Omega_3$}};
\draw [blue] (-2.0,-0.75) node  {{\large  $\Omega_4$}};
\draw [blue] (2.0,-0.75) node  {{\large  $\Omega_7$}};
\draw [blue] (1,1.75) node  {{\large  $\Omega_1$}};
\draw [blue] (1,-1.75) node  {{\large  $\Omega_6$}};
\draw [blue] (-1,1.75) node  {{\large  $\Omega_2$}};
\draw [blue] (-1,-1.75) node  {{\large  $\Omega_5$}};


\end{tikzpicture}

\end{center}
\label{fig:2}
\caption{The continuous spectra of $5\times 5$ Lax operator with  complex-valued $J$  subject to a  reduction of type
1) or 3) in (\ref{eq:Z2-nl})}
\end{figure}
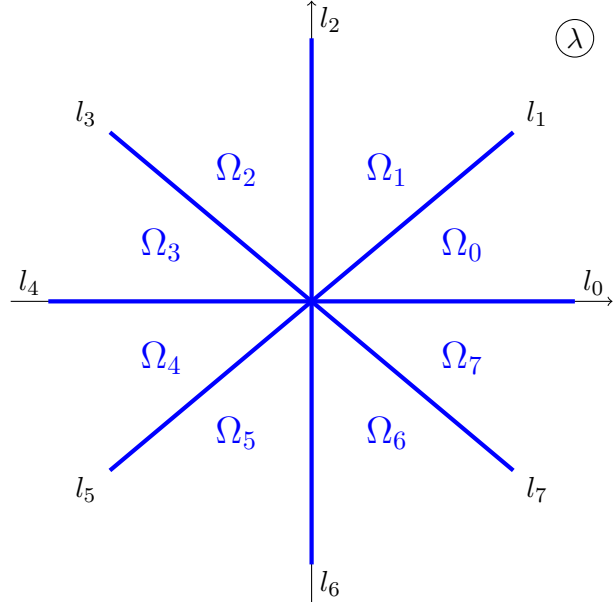

Therefore here we will consider mostly the case of complex-valued $J$; more specifically we will take $a_1=a_2^*$.
As in the previous Subsection we will take care that the $M$ operators has the same continuous spectrum as $L$,
which means that $I=a_2H_{e_1}+a_1 H_{e_2}= J^*$.

The construction of the FAS is quite analogous to the one for $sl(3)$ already describe above, see \cite{VYa,121}.
Skipping the details we note that the continuous spectrum of $L$ is located on the lines
\begin{equation}\label{eq:CBC}
{\rm Im}\, \lambda (\alpha, \vec{J}) =0, \qquad \alpha \in \Delta.
\end{equation}
This is a set of four lines intersecting at the origin, see Figure \ref{fig:2}. With each of the lines $l_\nu$ one can relate a root $\alpha$ (see Table
2) for which $  \im \, \lambda (\alpha, \vec{J})=0$ for $\lambda \in l_\nu$.

\begin{table}\label{tab:2}
\begin{center}
\begin{tabular}{|l|c|c|c|c|c|c|c|c|}
  \hline
  Ray & $l_0$ & $l_1$ & $l_2$ & $l_3$ & $l_4$ & $l_5$ & $l_6$ & $l_7$ \\\hline
  Root & $e_1+e_2$ & $e_2$ & $e_1-e_2$ & $e_1$ & $-(e_1+e_2)$ & $-e_2$ & $-(e_1-e_2)$ & $-e_1$ \\
  Angle & $0$ & $\varphi_1$ & $\pi/2$ & $\pi -\varphi_1$ & $\pi$ & $\pi+\varphi_1$ & $-\pi/2$ & $-\varphi_1$ \\
  \hline
\end{tabular}
\end{center}
\caption{To each of the rays $l_\nu$ of the continuous spectrum of the $so(5)$ Lax determined by (\ref{eq:CBC})
one can relate a root of the $so(5)$ algebra and the angle that it closes with the real axis, as shown above (here $\varphi_1= {\rm arg}\, a_1$).}
\end{table}

Quite analogously to the previous case, one can prove as particular case of the results in \cite{VYa,121} that:
\begin{equation}\label{eq:chi-nu3}\begin{aligned}
\chi_{\nu} (x,t, \lambda) &= \phi(x,t,\lambda) S^+_\nu (\lambda,t)=   \psi(x,t,\lambda) T^-_\nu (\lambda,t) D^+_\nu (\lambda), &\quad \lambda &\in l_\nu^+ ,\\
\chi_{\nu} (x,t, \lambda) &= \phi(x,t,\lambda) S^-_{\nu+1} (\lambda,t)=   \psi(x,t,\lambda) T^+_{\nu+1} (\lambda,t) D^-_{\nu+1} (\lambda), &\quad \lambda &\in l_{\nu+1} ^-,
\end{aligned}\end{equation}
where $\nu =0,\dots 7$ and
\begin{equation}\label{eq:chi-nu4}\begin{aligned}
\tilde{\chi}_{\nu} (x,t, \lambda) &= \phi(x,t,\lambda) S^+_\nu (\lambda,t) \hat{D}^+_\nu (\lambda)=   \psi(x,t,\lambda) T^-_\nu (\lambda,t) , &\quad \lambda &\in l_\nu^+ ,\\
\tilde{\chi}_{\nu} (x,t, \lambda) &= \phi(x,t,\lambda) S^-_{\nu+1} (\lambda,t)\hat{D}^-_{\nu+1} (\lambda)=   \psi(x,t,\lambda) T^+_{\nu+1} (\lambda,t) ,
&\quad \lambda &\in l_{\nu+1}^-,
\end{aligned}\end{equation}
 The scattering matrix $T_\nu(\lambda,t)$ for $\lambda\in l_\nu$ takes values in the $SL_\nu(2)$ subgroup of $SO(5)$ generated by
$E_{\alpha_\nu}$, $E_{-\alpha_\nu}$ and $H_{\alpha_\nu}$. By $S_\nu^\pm(\lambda)$,  $T_\nu^\pm(\lambda)$ and  $D_\nu^\pm(\lambda)$
we have denoted the Gauss factors of $T_\nu(\lambda,t)$ (see eq. (\ref{eq:ScatM}) above) and
\begin{equation}\label{eq:STDpm'}\begin{aligned}
S_\nu^\pm (\lambda) &= \exp \left( s_{\nu,\alpha_\nu}^\pm(\lambda) E_{\pm \alpha_\nu}\right), &\;
T_\nu^\pm (\lambda) &= \exp \left( t_{\nu,\alpha_\nu}^\pm (\lambda) E_{\pm \alpha_\nu}\right), &\;
D_\nu^\pm (\lambda) &= \exp \left( d_{\nu,\alpha_\nu}^\pm (\lambda) H_{\pm \alpha_\nu}\right),
\end{aligned}\end{equation}
By $\lambda \in l_\nu^\pm$ above we mean that $\lambda = \lambda' e^{\pm i0}$, where $\lambda'\in l_\nu$.

The solutions $\xi^\pm (x,t,\lambda) $ can be viewed also as solutions to a RHP
\begin{equation}\label{eq:RHPso5}\begin{split}
 \xi^+ (x,t,\lambda) = \xi^- (x,t,\lambda) G(x,t,\lambda), \qquad  G(x,t,\lambda) = e^{i\lambda Jx} G_0(t,\lambda)
 e^{-i\lambda Jx}
\end{split}\end{equation}
where $\nu =0,\dots 7$ with canonical normalization, i,e, $\lim_{\lambda \to\infty} \xi^\pm (x,t,\lambda) =\openone$.

Reducing the inverse scattering problem to a RHP first of all allows one to use the dressing method
for constructing the soliton solutions of the relevant $4$-wave equations.

\section{Dressing method and soliton solutions}\label{sec:4}

The dressing factor relates the FAS of two Lax operators:
\begin{equation}\label{eq:g0}\begin{split}
\chi_\nu^{(1)} (x,t,\lambda) = g(x,t,\lambda) \chi_\nu^{(0)} (x,t,\lambda),
\end{split}\end{equation}
where $\chi_\nu^{(1)} (x,t,\lambda)$ and $\chi_\nu^{(0)} (x,t,\lambda)$ are Lax operators with  potentials
$Q_1$ and $Q_0$. We assume that $Q_0=0$, so the corresponding `naked' solution is $\chi_\nu^{(0)} (x,t,\lambda)
= e^{-i\lambda Jx}$. Obviously the dressing factor must satisfy:
\begin{equation} \label{dress2}
 i g_x + [J,Q] g - g [J,Q_0] - \lambda [J,g]=0,
\end{equation}
Besides, $q(x,t,\lambda)$ must satisfy all the reduction conditions imposed on the Lax operator:
\begin{equation}\label{eq:g_reduction}\begin{aligned}
&\mbox{A)} & \quad s_0 g^\dag (-x,-t,-\lambda^*) s_0^{-1} &= \hat{g}(x,t,\lambda), \\
&\mbox{B)} & \quad s_0 g^\dag (-x,-t,\lambda^*) s_0^{-1} &= \hat{g}(x,t,\lambda) .
\end{aligned}\end{equation}
We will present below the 1- and 2- soliton solutions for nonlocal reductions of class $A)$.  The solutions of class B) are derived
analogously.

More about the dressing method the reader can find in \cite{ZMNP,GKKV*08,RI,Sh}.

\subsection{One soliton solution}\label{ssec:4.1.1}

Assume that for ${\frak g}\simeq sl(3,{\Bbb C})$ the dressing factor $g(x,t,\lambda)$ and its inverse  $\hat{g}(x,t,\lambda)$
have simple poles at $\lambda=\lambda_1^-$ and $\lambda=\lambda_1^+$ respectively, i.e.
\begin{equation} \label{dress1p}
 g(x,t,\lambda)= 1+\frac{\lambda_1^- - \lambda^+_1}{\lambda- \lambda_1^-}P_1(x,t), \qquad
 \hat{g}(x,t,\lambda)= 1+\frac{\lambda_1^+ - \lambda^-_1}{\lambda- \lambda_1^+}P_1(x,t),
\end{equation}
where $P_1(x,t)$ is the rank 1 projector:
\begin{equation}\label{eq:P1}\begin{split}
 P_1(x,t) = \frac{|n_0(x,t)\rangle \langle m_0(x,t)| }{\langle m_0(x,t)|n_0(x,t)\rangle}, \qquad P_1^2(x,t) = P_1(x,t).
\end{split}\end{equation}
From \eqref{eq:g_reduction} it follows that
\begin{equation} \label{dress1a}
g(x,t,\lambda)s_0g^{\dag}(-x,-t, -\lambda^* )s_0^{-1}=\openone,
\end{equation}
which means that:
\begin{equation} \label{P}
 P_1(x,t)=s_0P_1^{\dag} (-x,-t)s_0^{-1}, \qquad \lambda_1^+ =-(\lambda_1^-)^* = \alpha_1+i\beta_1.
\end{equation}
From the general theory of the dressing method \cite{ZaSh*74a} it is known that $q(x,t,\lambda)$ must satisfy the
equation
\begin{equation}\label{eq:dgdx}\begin{split}
i \frac{ dg}{dx} + ([J,Q_1] -\lambda J)g + g \lambda J =0, \qquad
i \frac{ dg}{dt} + ([J^*,Q_1] -\lambda J^*)g + g \lambda J^* =0,
\end{split}\end{equation}
where $Q_1(x,t)$ is the `dressed' potential determined by the one-soliton solution. In particular, analyzing the residue
for $\lambda \to \lambda_1^-$ as well as the limit for $\lambda\to\infty$ of this equation
there follows that
\begin{equation}\label{eq:dm0xt}\begin{split}
i \frac{ d \langle m_0|}{dx}  +\lambda_1^-  \langle m_0(x,t)| J &=0, \qquad i \frac{ d \langle m_0|}{dt}  +\lambda_1^-  \langle m_0(x,t)| J^* =0, \\
[J,Q_1(x,t) ]&= -2\alpha_1 [J, P_1(x,t)].
\end{split}\end{equation}
Thus we get that $m_{0k}(x,t) = m_{00k}e^{R_k+i\Omega_k}$, $n_{0k}(x,t) =(-1)^{k+1} m_{00 \bar{k}}^* e^{-R_{\bar{k}}+i\Omega_{\bar{k}}}$, $\bar{k}=4-k$
where
\begin{equation}\label{eq:mokxt}\begin{aligned}
R_1&= (\alpha_1j_1-\beta_1j_0) x - (\alpha_1j_1+\beta_1j_0) t, &\; \Omega_1&= -(\alpha_1j_0+\beta_1j_1) x - (\alpha_1j_0-\beta_1j_1) t, \\
R_2&= 2 \beta_1j_0( x+t), &\; \Omega_2&= 2\alpha_1j_0( x+t), \\
R_3&= -(\alpha_1j_1+\beta_1j_0) x + (\alpha_1j_1-\beta_1j_0) t, &\; \Omega_3&= -(\alpha_1j_0-\beta_1j_1) x - (\alpha_1j_0+\beta_1j_1) t,
\end{aligned}\end{equation}
For simplicity we also assume that $m_{003}=m_{001}^*$. Then for the denominator of $P_1(x,t)$ we get:
\begin{equation}\label{eq:Del1}\begin{split}
\langle n_0(x,t) | m_0(x,t)\rangle = e^{-i\Omega_2} \Delta_1(x,t), \quad
\Delta_1(x,t) =  2|m_{001}|^2 \cosh (R_1-R_3) - |m_{002}|^2 e^{3i\Omega_2}.
\end{split}\end{equation}
Finally, for the one-soliton solution we get:
\begin{equation}\label{eq:Q1s}\begin{aligned}
Q_{1;12}(x,t) &= -\frac{ 2\alpha_1}{\Delta_1} m_{001} m_{002} e^{R_2-R_3} e^{i(\Omega_2-\Omega_1)}, &\;
Q_{1;21}(x,t) &= \frac{ 2\alpha_1}{\Delta_1}  m_{001} m_{002}^* e^{R_1-R_2} e^{i(\Omega_2-\Omega_3)},\\
Q_{1;13}(x,t) &= -\frac{ 2\alpha_1}{\Delta_1} |m_{001}|^2  e^{i(\Omega_3-\Omega_1)}, &\;
Q_{1;31}(x,t) &= -\frac{ 2\alpha_1}{\Delta_1}  |m_{001}|^2  e^{i(\Omega_1-\Omega_3)},\\
Q_{1;23}(x,t) &= \frac{ 2\alpha_1}{\Delta_1} m_{001}^* m_{002}^* e^{R_3-R_2} e^{i(\Omega_2-\Omega_1)}, &\;
Q_{1;32}(x,t) &= -\frac{ 2\alpha_1}{\Delta_1}  m_{001}^* m_{002} e^{R_2-R_1} e^{i(\Omega_2-\Omega_3)}.
\end{aligned}\end{equation}

If we  assume  in addition $2|m_{01}|^2 >|m_{02}|^2$ then the denominator
{\em never vanishes} and {\em the soliton is regular.}

It is easy to check that for $x \to \pm \infty$ the solution decays to zero, i.e. it falls into the class of
potentials for which we constructed the FAS.

\subsection{Two soliton solution}

Let us now consider the two-soliton solution corresponds to two discrete eigenvalues, $\lambda_1$ and $\lambda_2$ so that the anzatz for $g$, becomes
\begin{equation}\label{eq:g2factor}\begin{split}
g(x,t,\lambda)&=\openone +\frac{A_1(x,t)}{\lambda-\lambda_1^-}+\frac{A_2(x,t) }{\lambda-\lambda_2^-}
\end{split}\end{equation}
for some residues $A_1(x,t)$ and $A_2(x,t)$.  From the condition \eqref{dress1a}  there follows that $\hat{g}(x,t,\lambda)$ must
have simple poles at $\lambda_k^+ =- (\lambda_k^-)^*$. In addition it leads to the following system of equations for $A_1$, $A_2$
\begin{equation}\label{eq:A12eqa}\begin{split}
\left( \openone -\frac{A_1(x,t)}{\lambda_k^+ - \lambda_1^-} -\frac{A_2(x,t) }{\lambda_k^+- \lambda^-_2}  \right)s_0A^{\dag}_{k}(-x, -t)s_0^{-1}=0 , \qquad k=1,2
\end{split}\end{equation}
which are the conditions of the vanishing of the residue of \eqref{dress1a} at $-\lambda^+_k$ and $-\lambda^*_2$.

Now the dressing factor (\ref{eq:g2factor}) relates the FAS of the `naked' Lax operator with the one for $L_2$ whose
potential $Q_2$ is the 2-soliton solution of the 3-wave equation. As before we consider $A_k(x,t)$ as rank-1
degenerate matrices in the form
\begin{equation}\label{eq:A12nm}\begin{split}
 A_k(x,t)=|n_k(x,t)\rangle \langle m_k(x,t)|,
\end{split}\end{equation}
where $|n_k(x,t) \rangle $ are 3-component vector-columns, and  $\langle m_k(x,t)|$
 are 3-component vector-rows.  The system \eqref{eq:A12eqa} produces the following linear system of two equations
\begin{equation}\label{eq:1-2}\begin{split}
\left(\begin{array}{c} s_0 |m_1^*(-x,-t) \rangle \\  s_0 |m_2^*(-x,-t) \rangle  \end{array}\right) + \left(\begin{array}{cc} F_{11} & F_{12} \\
 F_{21} & F_{22} \end{array}\right) \left(\begin{array}{ccc} | n_1(x,t) \rangle \\ | n_2(x,t) \rangle   \end{array}\right)=0,
\end{split}\end{equation}
where
\begin{equation}\label{eq:Fkm}\begin{aligned}
F_{kp} = \frac{\langle m_p(x,t)|s_0 |m_k^*(-x,-t)\rangle }{\lambda_k^+ +\lambda_p^{+,*}}.
\end{aligned}\end{equation}
which allows the expression of the $ n$- vectors via $m$- vectors.
In addition, the condition \eqref{dress2} requires the residues to satisfy the equations
\begin{equation}\label{eq:ResDress}\begin{split}
 i \frac{dA_k }{dx}+[J,Q]A_k- \lambda_k^- [J, A_k]=0, \qquad
 i \frac{dA_k }{dt}+[J^*,Q]A_k- \lambda_k^- [J^*, A_k]=0,
\end{split}\end{equation}
$k=1,2.$
These differential equations for $A_i$ lead to the following  equations for  $|n \rangle $ and  $\langle m|$:
\begin{equation}\label{eq:dmkdx}\begin{aligned}
i \frac{ d \langle m_k|}{dx} + \lambda_k^- \langle m_k|  J &=0,  &\qquad
i \frac{d |n_k\rangle}{dx} + ([J,Q_2]- \lambda_k^- J)|n_k\rangle &=0,\\
i \frac{d \langle m_k|}{dt} + \lambda_k^- \langle m_k|  J^* &=0,  &\qquad
i \frac{d |n_k\rangle}{dt} + ([J,Q_2]- \lambda_k^- J^*)|n_k\rangle &=0.
\end{aligned}\end{equation}
This  result shows that while $\langle m_k |  $   satisfy the 'naked' equation
 at $\lambda=\lambda_k^-$ then $|n_k\rangle$  are eigenfunctions
of the spectral problem for $\lambda=\lambda_k^-$  for the (yet unknown) 2-soliton solution $Q_2(x,t)$.
One can easily solve for $|m_k \rangle $  and then can recover $|n_k\rangle$
from \eqref{eq:1-2}. These solutions are
\begin{equation}\label{eq:Palarisation}\begin{split}
& \langle m_k|=\langle m_{k0}|e^{i\lambda_k^- (Jx+J^*t)}, \\
\end{split}\end{equation}
Taking $\lambda_k^\pm = \pm \alpha_k + i\beta_k$ we find
\begin{equation}\label{eq:n1-2}\begin{split}
\left(\begin{array}{ccc} | n_1(x,t) \rangle \\ | n_2(x,t) \rangle   \end{array}\right) =
\frac{1}{\Delta_2(x,t)} \left(\begin{array}{cc} F_{22} & -F_{12} \\ - F_{21} & F_{11} \end{array}\right)
\left(\begin{array}{c} s_0 |m_1^*(-x,-t) \rangle \\  s_0 |m_2^*(-x,-t) \rangle  \end{array}\right) ,
\end{split}\end{equation}
where $ \Delta_2(x,t) = F_{12} F_{21} - F_{11} F_{22}$.
Clearly $\Delta_2(x,t)$ is $\mathcal{ PT}$ invariant. The 2-soliton solution is determined from
\begin{equation}\label{eq:Q2s}\begin{split}
[J, Q_2(x,t)= [J, A_1(x,t)+A_2(x,t)] =  \left[ J, |n_1(x,t) \rangle \langle m_1(x,t) | +|n_2(x,t) \rangle \langle m_2(x,t) | \right]
\end{split}\end{equation}
Thus the two-soliton solution will be  complicated rational functions of the exponentials like $e^{-R_{k}\pm i\Omega_{k}}$.
In the generic case it is rather difficult to find out if this solution is singular or not.

However for some special cases of the parameters the 2-soliton solutions  simplify and one can see that
they are {\em regular}. First we assume that $\lambda_1^+=-\lambda_2^+ =\alpha_1+i\beta_1$. That means that
\begin{equation}\label{eq:m12}\begin{aligned}
\langle m_1(x,t)| &= \langle \mu_0| e^{\mathbf{R} + i \mathbf{\Omega}}, &\quad
|m_1^*(-x,-t)\rangle &= e^{-\mathbf{R} + i \mathbf{\Omega}} |\mu_0^*\rangle, &\quad \mathbf{R} &= \diag( R_1,R_2,R_3) \\
\langle m_2(x,t)| &= \langle \tilde{\mu}_0| e^{-\mathbf{R} - i \mathbf{\Omega}},  &\quad
|m_2^*(-x,-t)\rangle &= e^{\mathbf{R} - i \mathbf{\Omega}} |\tilde{\mu}_0^*\rangle, &\quad
\mathbf{\Omega} &= \diag( \Omega_1, \Omega_2, \Omega_3)
\end{aligned}\end{equation}
Next, like for the one-soliton case above, we assume $\mu_{03}=\mu_{01}^*$ and $\tilde{\mu}_{03}=\tilde{\mu}_{01}^*$
and $\tilde{\mu}_{02}=0$.
Then the expressions for $F_{kp}$ simplify into:
\begin{equation}\label{eq:Fpk}\begin{aligned}
F_{11} &= \frac{e^{-i\Omega_2}}{\alpha_1}  \left( |\mu_{01}|^2 \cosh (R_1-R_3) - \frac{1}{2} |\nu_{02}|^2 e^{-3i\Omega_2} \right),
\quad F_{22} = -\frac{e^{i\Omega_2} }{\alpha_1}  |\mu_{01}|^2 \cosh (R_1-R_3) , \\
F_{12} &= \frac{e^{R_2}}{i\beta_1}  |\mu_{01}\tilde{\mu}_{01}| \cos (\Omega_1 -\Omega_3 -\sigma_{01}),
\qquad F_{21} = -\frac{e^{-R_2}}{i\beta_1}  |\mu_{01}\tilde{\mu}_{01}| \cos (\Omega_1 -\Omega_3 +\sigma_{01}).
\end{aligned}\end{equation}
where $\sigma_{01} = \arg \mu_{01} + \arg \tilde{\mu}_{01}$. Inserting these expressions into $\Delta_2$ we obtain:
\begin{multline}\label{eq:Del2}
\Delta_2(x,t) = \frac{1}{\alpha_1^2} |\tilde{\mu}_{01}|^2 \cosh(R_1-R_3) \left( |\mu_{01}^2 \cosh (R_1-R_3) -
\frac{1}{2} |\mu_{02}|^2 e^{-3i \Omega_2} \right) \\
+ \frac{1}{\beta_1^2} |\mu_{01}\tilde{\mu}_{01}|^2  \cos (\Omega_1 -\Omega_3 +\sigma_{01})  \cos (\Omega_1 -\Omega_3 -\sigma_{01}).
\end{multline}
The above expression will have no zeroes if, like in the one-soliton case above, we  assume that  $2|\mu_{01}| >|\mu_{02}$  and require
in addition $\sigma_{01}=0$ or $\sigma_{01}=\pi$
Then the corresponding two-soliton solutions will be regular.

\section{Conclusions}\label{sec:6}

We have constructed the Fundamental Analytic Solutions for nonlocal reductions of $3$- and $4$- wave
systems (related to the algebras $sl(3,{\Bbb C})$  and $so(5,{\Bbb C})$ respectively) having ${\cal PT}$-symmetry and
briefly outlined the spectral properties of the associated Lax operators. Furthermore, we have derived the one- and two-soliton solutions of
the 3-wave equations with complex $J$.

Analysis of the one soliton solutions shows that they can be regular for all $x$ and $t$
provided the polarization vectors satisfy certain constraints, see Subsection 5.1.
We also find specific set of parameters for the
two-soliton solutions of the $\mathcal{PT}$ symmetric $3$-wave equations  that ensure their regularity.

These examples demonstrate that the $\mathcal{PT}$ symmetric $3$-wave equations may have
 regular multi-soliton solutions for some specific choices of their parameters. Similar analysis can be applied also
 to the $4$-wave  $\mathcal{PT}$ symmetric  equations. Details about it will be published elsewhere.

\section*{Acknowledgements}
We thank an anonymous referee for careful reading the text and for useful remarks.

\end{document}